\documentclass[preprint,aps,superscriptaddress,floatfix,showpacs]{revtex4}
\usepackage{epsfig}
\usepackage{multirow}
\usepackage{rotating}
\usepackage{color}
\usepackage{algorithm}
\usepackage{algorithmic}
\floatstyle{plain}
\newfloat{mybox}{thp}{lp}
\floatname{mybox}{Algorithm}
\newcommand{\noop}[1]{}
\newcommand{\rc}[1]{{\color{black} #1}}
\newcommand{\MVM}{MSG-VM }
\newcommand{\MVMA}{MSG-VM}
\begin{document}
\title{Multistep greedy algorithm identifies community structure in real-world and computer-generated networks}
\author{Philipp Schuetz and Amedeo Caflisch}
\affiliation{Department of Biochemistry, University of Zurich, Winterthurerstrasse 190,
CH-8057 Zurich, Switzerland  \\ Fax: +41 44 635 68 62, Email: {caflisch}@bioc.uzh.ch}
\date{July 4, 2008}
\begin{abstract}
We have recently introduced a multistep extension of the greedy
algorithm for modularity optimization. The extension 
 is based on the idea that merging $l$ pairs of communities ($l>1$)
at each iteration prevents premature condensation into few large communities.
Here, an empirical formula is presented  for the choice of the step width $l$ that generates
partitions with (close to) optimal modularity for 17 real-world and 1100 computer-generated networks.
Furthermore, an in-depth analysis of the communities of
two real-world networks (the metabolic network of the bacterium {\em E.~coli} and the graph of coappearing words in the titles of papers coauthored by Martin Karplus)
provides evidence that the partition obtained by the multistep greedy algorithm
is superior to the one generated
by the original greedy algorithm not only with respect to  modularity but also
according to objective criteria.
In other words, the multistep extension of the greedy algorithm
reduces the danger of getting trapped in local optima of modularity
and generates more reasonable partitions.
\end{abstract}
\maketitle

\section{Introduction}
The coarse-grained organization of many real-world networks manifests itself in a natural divisibility of the vertices into modules (or communities). A community is a set of vertices that are more connected among each other than with vertices of other communities. Community structure has been reported for social networks \cite{Newman2001b,Gleiser2003}, metabolic networks \cite{Ma2003,Guimera2005,Muff2005}, and protein folding networks  \noop{\cite{Rao2004}\cite{Hubner2005}} \cite{Rao2004,Hubner2005,Gfeller2007,Muff2008,Krivov2008}. Several procedures have been developed to partition a network into modules. Often applied are techniques that rely on the optimization of a scoring function called \textit{modularity} \cite{Newman2004}. This assessment function compares the fraction of edges within a module with its expectation value in the case of randomly connected vertices with equal degree sequence. The modularity is defined as
\begin{equation}
Q = \sum_{i=1}^{N_C} \left [ \frac{I(i)}{L} - \left ( \frac{d_i}{2 L} \right )^2 \right ],   \label{modularity}
\end{equation}
with $I(i)$ being the weights of all edges linking vertices of community $i$, $d_i$ the sum over all vertex degrees in module  $i$, $L$ the total edge weight, and $N_C$ the number of communities. The optimization of modularity has been proven to be a NP-hard problem \cite{Brandes2006}. Thus, heuristic techniques such as extremal optimization \cite{Duch2005}, simulated annealing \cite{Guimera2005}, and the greedy algorithm \cite{Newman2004a} have been developed and applied to find partitions with high modularity. Because of the global character of modularity [i.e., in Eq.~(\ref{modularity}) the connectivity and degree of each community are compared with the edge weight of the whole network], it has been shown that modules smaller than a certain scale cannot be resolved \cite{Fortunato2007}. In other words, small communities are amalgamated with others instead of being detected autonomously.  A higher resolution variant of modularity, called \textit{localized} modularity, addresses the limit on the detectable community size \cite{Muff2005}.

Recently, we have introduced a multistep extension of the greedy algorithm (MSG) and combined it with a simple vertex-by-vertex refinement procedure [vertex mover VM] \cite{Schuetz2007}. The essential idea of the MSG algorithm is to promote the simultaneous merging of several pairs of communities to prevent premature trapping in a local optimum of modularity. Given an appropriate choice of the step width $l$, the \MVM~algorithm finds partitions with high modularity in \rc{short running time. Our implementation of the \MVM algorithm \cite{Schuetz2007,Karplus} has the same scaling behavior as the efficient version of the greedy algorithm \cite{Clauset2004}, which has the smallest complexity among the commonly used community-detection algorithms \cite{Danon2005}. Note that the running time of both the \MVM algorithm \cite{Schuetz2007} and the greedy algorithm \cite{Clauset2004} is $O(D L \log N)$ with $L$, $N$, and $D$ the number of edges, vertices, and the depth of the dendrogram describing the community structure, respectively. For a sparse network with $L\sim N$ and $D\sim \log N$, the scaling is essentially linear $O(N \; \log^2 N)$.}

In this paper, we derive an empirical formula for predicting optimal $l$ values, i.e., values of the step width that yield a modularity very close to the  highest value achievable by the \MVM~algorithm. Furthermore, for two real-world networks having each an inherent partition into substructures, we compare the community structures identified by the original greedy and the \MVM algorithm. These real-world examples are the metabolic network of \textit{E.~coli} and the graph of coappearing words in the titles of publications coauthored by Martin Karplus, the most cited theoretical chemist. The inherent substructures of the former are the metabolic pathways, while the inherent substructure of the network of Karplus' paper titles are the sets of words predominantly used in research subfields in theoretical and computational chemistry. These two examples illustrate that the \MVM algorithm detects the underlying substructures more accurately than the original greedy algorithm.

\section{Methods} \label{Methods}
\subsection{Multistep greedy and vertex mover algorithms}
The MSG algorithm optimizes modularity by an iterative procedure in which multiple pairs of communities are merged at each iteration. This multistep approach is a significant extension with respect to the original greedy algorithm \cite{Newman2004a}, in which only the pair of communities that improves modularity most is merged in each iteration. A pseudocode description of the MSG algorithm is given below (see Algorithm \ref{concept}). Note that the step width $l$ influences the number of merged pairs (but is not necessarily identical to it); furthermore, $l$ is kept constant during an MSG run (for more details, the reader is referred to the original publication \cite{Schuetz2007}).

\begin{mybox}
 \hrule \vspace*{3pt}
\footnotesize
 \begin{algorithmic}
\STATE \textit{Initialization:}
\STATE Each vertex is a community;
\STATE  Calculate matrix  $\Delta Q$ whose elements are the modularity changes upon merging of module pair $(i,j)$;
\STATE \textit{Iteration:}
\WHILE{pair $(i,j)$ with $\Delta Q_{ij}> 0$ exists}
\FORALL{triplets $(i,j,\Delta Q_{ij})$ of $\Delta Q$, parsed w.r.t. decreasing $\Delta Q_{ij}$  and increasing $ (i,j)$}
\IF{ $\left \{ \begin{array}{l}
                \Delta Q_{ij} > 0 \mathrm{\;in \; best \;} l  \; \mathrm{values \;  in\; }\Delta Q \mathrm{-matrix}\\
 i \mathrm{\; and \;} j\mathrm{\; unchanged\; in \; iteration} \\
 
               \end{array} \right \}$}
\STATE \vspace*{2mm}\hspace*{2.1cm}MergeCommunities(i,j);
\ENDIF
\ENDFOR
\ENDWHILE
 \end{algorithmic}
\hrule
\caption{Flowchart of the MSG procedure. Details of the efficient merge of two communities and the calculation of the modularity change matrix are presented in \cite{Schuetz2007}. \label{concept}}
\end{mybox}

Applied upon convergence of the MSG algorithm the VM procedure improves modularity by ``adjusting'' misplaced vertices. The VM procedure parses the vertex list in ascending vertex degree and index order  and checks for each vertex whether a reassignment to one of the neighboring communities yields a modularity improvement \cite{Schuetz2007}. 


\subsection{Networks} \label{networks}
All networks in this article are treated undirected and unweighted.
\begin{table*}
\scalebox{0.84}{
\begin{tabular}{llrrcccccrccc}
& & & & &  \multicolumn{2}{c}{\parbox{2.6cm}{\MVM with Optimal $l$}} & &  \multicolumn{1}{c}{\parbox{2.6cm}{\MVM with $l$ from Eq.~(\ref{prediction})}} & &  \multicolumn{2}{c}{\parbox{2.6cm}{\MVM with Random $l$}}\\ \cline{6-7}\cline{9-9} \cline{11-12}
Network  &  Ref. & Vertices & \multicolumn{1}{c}{\phantom{i}Edges ($L$)\phantom{i}} & \phantom{MMi} & $l_{\rm opt}/\sqrt{L}$ & $Q_{\rm opt}$   &    &  \phantom{i}$Q_{\rm pred}$\phantom{i}  &    &  \phantom{M}$\langle Q_{\rm rand} \rangle$\phantom{M} & $\langle Q_{\rm rand}^{l< 1.5 \sqrt{L}} \rangle$\\ \hline 
Zachary Karate Club &   \cite{Zachary1977}    &  34  &  78 &   &  0.34&  0.398   &    &  0.398   &  &  0.391  &  0.398 \\
Metabolic \textit{E.~coli}  &   \cite{Ma2003}   &  443  &  586 &   &  0.25 &  0.816   &   &  0.816    &  &  0.813  &  0.816\\
College Football &   \cite{Girvan2002}   &  115  &  613 &   &  0.04&  0.603   &  &  0.595  &   &0.579  & \it 0.596 \\
Metabolic \textit{C.~elegans} &   \cite{Jeong2000}  &  453  &  1899 &   &  4.80 &  0.450   &  &  0.447   &  & 0.439  &  0.445\\
Jazz   &   \cite{Gleiser2003}   &  198  &  2742 &   &  10.81  &  0.4451     &   &  0.4447   &   & \it 0.4451 & \it 0.4448\\
Email   &   \cite{Guimera2003}   &  1133  &  5451 &   &  0.76 &  0.575    &  &  0.575  &   &  0.564  &  0.574 \\
Yeast  (PPI, LCC)  &   \cite{Krogan2006}   &  2552  &  7031 &   &  0.42 &  0.706   & &  0.705  &   &  0.693  &  0.702 \\
M.~Karplus & \cite{Schuetz2007,Karplus}     &  1167  &  13423 &   &  0.79 &  0.316    &  &  0.311  &   &  0.306  &  0.311\\
PPI  \textit{S.~cerevisiae} (LCC)&   \cite{Colizza2005}   &  4626  &  14801&    &  1.40  &  0.545   &   &  0.544   &  &0.531  &  0.543 \\
PPI \textit{S.~cerevisiae}  &   \cite{Colizza2005}   &  4713  &  14846 &   &  1.40&  0.546   &  &  0.546   &  & 0.532  &  0.545 \\
Internet    &   \cite{AS2001}   &  11174  &  23409 &   &  1.82 &  0.625    &  &  0.619   &  &0.615  &  0.618 \\
PGP-key signing  &   \cite{Guardiola2002,Boguna2004}  &  10680  &  24340  &  &  0.28 &  0.878    &   &  0.876   &  &0.873  &  0.876\\
Word Association (LCC)&   \cite{Nelson2004}   &  7204  &  31783  &   &  0.40 &  0.541   &  &  0.536  &    &   0.528  &  0.536\\
Word Association &   \cite{Nelson2004}    &  7207  &  31784  &   &  0.54&  0.540     &  &  0.537  &  &  0.527  &  0.536\\
Collaboration  &  \cite{Newman2001b}   &  27519  &  116181  &   &  0.45  &  0.748    &  &  0.746  &   &  0.743  &  0.744 \\
WWW  &   \cite{Albert1999}   &  325729  &  1117563  &   &  2.87 &  0.939   &    &  0.936  &   & \it 0.937  & \it 0.937 \\
Actor  &   \cite{Barabasi1999}   &  82583  &  3666738  &  &  1.27    &  0.543    & &  0.536  &   & \it 0.537  & \it 0.539\\ \hline
\end{tabular}}
\caption{\label{real-world-basic} \label{real-world-estimation}Properties of real-world networks and comparison of \MVM runs using $l$ as in Eq.~(\ref{prediction}) or picked at random. The column ``$Q_{\rm opt}$'' lists the maximal value of modularity obtained by running \MVM for all values of $l$ smaller than $\min\{5000,L\}$ (where $L$ is the number of edges). The column ``$Q_{\rm pred}$'' lists the \MVM modularity obtained using Eq.~(\ref{prediction}) to determine the step width. The columns ``$\langle Q_{\rm rand} \rangle$'' and ``$\langle Q_{\rm rand}^{l< 1.5 \sqrt{L}} \rangle$'' show the expectation value for the \MVM modularity when six values of $l$ are picked randomly from a uniform distribution in the range $1 \leq l \leq \min\{5000,L\}$ and $1 \leq l \leq 1.5 \sqrt{L}$, respectively. The expectation value is estimated by averaging, over 1000 samples, the highest modularity obtained using six values of $l$ (details are given in Sec.~VII of the Supplementary Material \cite{suppl-mat}). Six values of $l$ are picked  randomly for each sample because six values were used to determine $Q_{\rm pred}$: the four values of $l$ calculated by Eq.~(\ref{prediction}) and the two integers adjacent to the best of these four. Values of $\langle Q_{\rm rand} \rangle$ and $\langle Q_{\rm rand}^{l<1.5 \sqrt{L}}\rangle$ higher than the corresponding $Q_{\rm pred}$ are in italics. The acronym LCC stands for ``largest connected component''. }
\end{table*}

\subsubsection{Real-world networks}
The real-world networks are the same as in \cite{Schuetz2007} and are listed in Table \ref{real-world-basic}. Sociological applications are included with the Zachary karate club example \cite{Zachary1977}, the conference graph of college football teams \cite{Girvan2002}, the graph of jazz groups with common musicians \cite{Gleiser2003}, the network of mutual trust (PGP-key signing) \cite{Guardiola2002,Boguna2004}, the  collaboration network (coauthorships in cond-mat articles) \cite{Newman2001b} and the graph of costarring actors in the IMDB database \cite{Barabasi1999}. Network applications in biochemistry are covered by the graph of metabolic reactions in the nematode \textit{Caenorhabditis elegans} \cite{Jeong2000} and the bacterium \textit{Escherichia coli} \cite{Ma2003} as well as two different data sets describing the protein-protein interactions in \textit{Saccharomyces cerevisiae} (budding yeast) \cite{Krogan2006,Colizza2005} with labels ``PPI'' and ``yeast''. Linguistic applications are covered by the Word Association network \cite{Nelson2004} and the graph of the coappearing words in titles of publications (co)authored by Martin Karplus \cite{Schuetz2007,Karplus} who has the third highest $h$ factor \cite{Hirsch2005} among chemists  \cite{Ball2007}. From computer science the internet routing network \cite{AS2001} and the graph of WWW pages \cite{Albert1999} are included. The effects of disconnected graphs  are considered by including the full network as well as its largest connected component (LCC). 

\subsubsection{Computer-generated networks} \label{computerGenerated}
\begin{table}
\begin{tabular}{lccccclccccc}  \hline
 Type & \phantom{i}\parbox{2cm}{\setlength{\baselineskip}{12pt} No.~of realizations \vspace*{-3mm}} \phantom{i}&\hspace*{3mm}& \multicolumn{1}{c}{Vertices} & \hspace*{3mm}&\multicolumn{1}{c}{Edges} & \hspace*{3mm}& Remarks \\ \hline
GN$_1$ & 100 & & \multicolumn{1}{c}{128} & & \multicolumn{1}{c}{1024} & & $z_{\rm out}=3$ - $16$  \\
GN$_2$ &100 & & \multicolumn{1}{c}{128} &  & \multicolumn{1}{c}{512} & &  $z_{\rm out}=2$ - $8$  \\
GN$_3$ &100 & & \multicolumn{1}{c}{128} & &  \multicolumn{1}{c}{2048} & &  $z_{\rm  out}=2$ - $32$ \\[2mm]
SED & 300 & & 11-976 &  & 10-19247 &  & Exp. deg. distr. \\ 
SLD &200 & & 19-3777 &  & 43-78741 &  & Linear deg. distr. \phantom{i}  \\ 
LLD & 300 & & 309-4278 &  & 1523-342940&  & Linear deg. distr.  \\
\hline 
\end{tabular}
\caption{\label{random} \label{results-random}Properties of computer-generated networks. The networks in the GN$_{i}$ (Girvan and Newman) sets ($i=1,2,3$) possess an imposed four community structure where $z_{\rm out}$ controls the average number of edges connecting two different modules \cite{Girvan2002}. For the networks of type SED (small networks with exponential degree distribution), SLD (small networks with linear degree distribution), and LLD (large networks with linear degree distribution) a degree distribution has been prescribed to foster the formation of communities.}
\end{table}

\rc{A total of 1100 computer-generated networks were used for an in-depth assessment of the empirical formula for the prediction of optimal values of $l$} (Table \ref{random}). The networks in GN$_{1,2,3}$ consist of 128 vertices organized in four equally sized communities \cite{Girvan2002} . The cohesion of the vertices within a module is controlled by a parameter called $z_{\rm out}$ which determines the average number of edges connecting vertices of different modules. To consider clearly formed/loosly coupled modules the $z_{\rm out}$ parameter is chosen uniformly from the second smallest to the highest value. Among the sets GN$_{1,2,3}$, the number of edges is varied to assess the effect of different values of average degree. 

The remaining \rc{test cases are larger networks, which have no imposed community structure and a heterogeneous distribution of the vertex degrees and community sizes (confer Table 1 in the supplementary material \cite{suppl-mat}). A recent study, published after the submission of this work, has emphasized the importance of this heterogeneity for testing community-detection algorithms on severe benchmarks \cite{Lancichinetti2008}}. To foster a ``spontaneous'' formation of modules a vertex degree distribution is imposed. The network is generated by choosing a number of vertices at random (uniform distribution), assigning edge endpoints to each vertex according to the degree distribution and joining the edge endpoints at random. To examine the effect of different degree distributions, an exponential distribution is used for the networks in SED (small networks with exponential degree distribution) and a linear distribution is imposed on the networks in SLD and LLD. All networks in LLD have at least 300 vertices. After generation, the networks in SED, SLD and LLD are projected onto the biggest connected component and treated as unweighted.

\section{Results}\label{benchmarking-procedure}
It is helpful  to recall here that $L$ is the number of edges and $l_{\rm opt}$ is the value of the step width that yields the highest \MVM modularity (among all tested values of step width). The \MVM~algorithm is applied on each real-world network using every integer $l < \min\{5000,L\}$. The modularity values before and after the VM application are recorded. For the computer-generated networks all integer values $l < 10 \sqrt{L}$ have been tested (the $\sqrt{L}$ scaling is rationalized in the next subsection).
\subsection{Dependence of $l$ on network properties} \label{expected-scaling}
The correlation between the optimal step width $l_{\rm opt}$ and several topological properties was calculated. The following properties or powers thereof were used: number of vertices and edges, highest degree, average degree, standard deviation of degree, average of power 1, 2, or 3 of the clustering coefficient, and average and standard deviation of the differences in clustering coefficient values or degree of linked vertices. The highest correlation was observed for $\sqrt{L}$ (0.7728, correlation coefficients of other properties are listed in the supplementary material \cite{suppl-mat}).

This empirical result is consistent with the $\sqrt{L}$ dependence of the number of communities yielding maximal modularity as recently demonstrated for one class of networks \cite{Fortunato2007}. In fact, a close inspection of the MSG algorithm shows that the step width $l$ determines the number of communities formed during the first 1\% - 5\% of the iterations (the number of iterations is strongly dependent on the network topology). Each module in the final solution has to be nucleated as early as possible and therefore $l$ to be chosen according to the expected number of communities. 

\begin{figure*}
 \centering
 \includegraphics[width=12cm]{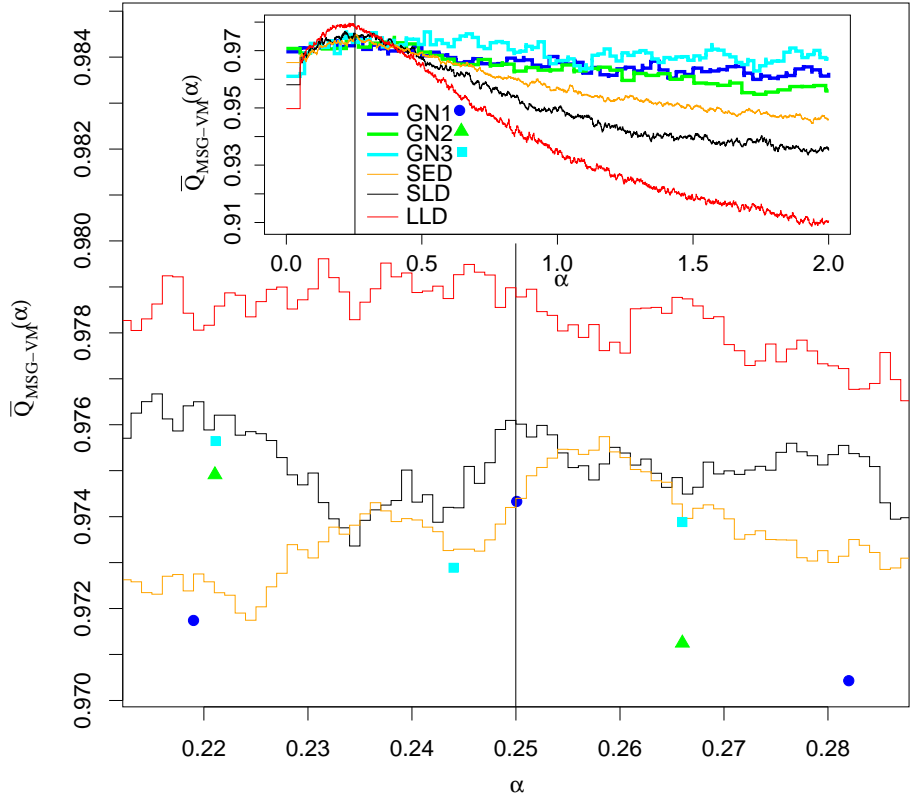}
 \caption{\label{performance-profiles}(Color online) Dependence of $Q_{\mbox{\tiny \MVM}}$ on the $\sqrt{L}$ prefactor $\alpha$ for the computer-generated networks. The averages are taken separately for each type of computer-generated networks.  The vertical black line denotes $\alpha=0.25$, which is the value suggested in Eq.~(\ref{prediction}). The parameter range for $\alpha$ has been discretized to multiples of 0.001 to simplify the calculations.} 
\end{figure*}

\subsubsection{Optimal prefactor for computer-generated networks} \label{average}
To determine the prefactor $\alpha$ in the $\sqrt{L}$-scaling law the computer-generated networks introduced in Sec.~\ref{computerGenerated} are examined first. This choice is due to their multitude (1100 networks) and their lack of overlapping condensed structures [i.e., few (almost) complete subgraphs sharing vertices] as consequence of the construction principle. First, we observe that for 97 of the 1100 networks the \MVM modularity does not depend on $l$. Further, for each value of $\alpha$ the \MVM modularity is averaged over all networks of the same type $\bar{Q}_{\mbox{\tiny \MVMA}}(\alpha) = \frac{1}{N_S} \sum_{i \in  S} \frac{Q_{\mbox{\tiny \MVMA}}^i\left (\left \lfloor \alpha \sqrt{L_i}\right \rfloor \right ) }{\max _l\left ( Q_{\mbox{\tiny\MVMA}}^i(l)\right ) }$, where $S$ is the type of networks, $N_S$ is the number of networks of type $S$,  $\lfloor . \rfloor$ is the floor function, and $L_i$ is the number of edges in network $i$. All $\alpha$ profiles peak for $0.2 < \alpha < 0.3$  and show a similar behavior (Fig.~\ref{performance-profiles}). The $\alpha$ profiles averaged over all computer-generated networks peak at $\alpha=0.251$. [It is legitimate to consider the average because for each $\alpha$ the histogram of $\frac{Q_{\mbox{\tiny \MVMA}}^i\left (\left \lfloor \alpha \sqrt{L_i}\right \rfloor \right ) }{\max _l\left ( Q_{\mbox{\tiny\MVMA}}^i(l)\right )}$ ($i$ indexing the network realizations) follows an unimodal distribution with an additional peak at 1.0 originating from the degeneracy of $l_{\rm opt}^i$.] Excluding the additional peak, the highest normalized modularities are still observed for $0.2 < \alpha < 0.3$.  Remarkably, the degeneracy of $l_{\rm opt}^i$ [i.e., the number of networks with $Q_{\mbox{\tiny \MVMA}}^i(\lfloor \alpha \sqrt{L_i}\rfloor) = \max_l\left (Q_{\mbox{\tiny \MVMA}}^i(l) \right )$] is highest for $0.18 < \alpha < 0.26$. A leave-$N$-out procedure (confer supplementary material \cite{suppl-mat} for details) provides evidence that $\alpha=0.251$ would have been (close to) optimal also for another selection of networks. 
The application of the \MVM algorithm with step width $\lfloor 0.251 \sqrt{L}\rfloor$ yields 97.6\% of the highest \MVM modularity averaging over all computer-generated networks (98\% if median is calculated).

\subsubsection{Comparison of empirical formula with random selection of step width}
If a step width value is selected at random among $l < \min\{L,5000\}$ (all tested values), the \MVM algorithm is expected to yield 93.4\% of the highest \MVM modularity on average over all computer-generated networks [the expectation value is equal to the arithmetic mean over all $Q_{\mbox{\MVM}}(l)$ values]. An in-depth analysis (details given in the supplementary material \cite{suppl-mat}) shows that $l_{\rm opt} < 1.5 \sqrt{L}$ for 92.6\% of all computer-generated networks. If a step width value smaller than $1.5 \sqrt{L}$ is chosen at random, the expectation value of the \MVM modularity raises to 95.9\% of its highest value (average over all computer-generated networks). Thus, the empirical formula $l = 0.251 \sqrt{L}$ performs 4.3\% better (of a maximum of 6.6\%) than a value of step width picked at random if all tested values are considered. If the reduced test set $l < 1.5 \sqrt{L}$ is used, the empirical formula performs 1.7\% better than a value of step width picked at random (4.1\% maximal improvement). More precisely, for 85.5\% of the networks the \MVM modularity with $l = 0.251 \sqrt{L}$ is higher than the one with $l$ picked at random and the average improvement for these networks is 2.4\%. 

To account for limited sampling the prefactor $\alpha=0.25$ is assumed to be optimal for the computer-generated networks (the prefactors 0.251 and $0.25$ can be considered identical as the real to integer conversion yields the same value of $l$ for networks with $L<10^6$). 

\subsubsection{Application to real-world networks}

In comparison to computer-generated graphs, real-world networks are endowed with more condensed substructures. Therefore, a different scaling behavior than for the computer-generated networks is possible. To improve statistics and reduce spurious effects due to vertex labeling artifacts (a value of step width yields a high \MVM modularity as it profits exclusively from the ``right'' parsing of the vertices), 100 copies of the smallest 10 real-world networks are created with permuted vertex labelings (details are presented in the supplementary material \cite{suppl-mat}). For each copy the influence of $l$ is tested as described in Sec.~\ref{benchmarking-procedure}. Except for the College Football and Email networks all $\bar{Q}_{\mbox{\tiny \MVM}}$ profiles (confer Sec.~\ref{average} for the definition) averaged over the scrambled variants are observed to peak for values of step width equal or very close to 
\begin{equation}\label{prediction}
l=\left \lfloor \alpha \;  \sqrt{L}\right \rfloor \quad (\alpha=0.25, 0.5, 0.75, 1) 
\end{equation}
(supplementary material \cite{suppl-mat}). The \MVM modularity deviates at most by 1.47\% from the maximal value (Table \ref{real-world-estimation}). Moreover,  for 13 of the 17 networks the deviation is smaller than 0.94\%. In comparison to the effect of permuted vertex labels  this deviation is of the same order of magnitude and thus negligible (details given in the supplementary material \cite{suppl-mat}). 

To further assess the predictive power of Eq.~(\ref{prediction}), the \MVM modularity obtained with $l$ as in Eq.~(\ref{prediction}) is compared with a random selection of the step widths. Because of the real to integer conversion induced by the floor function, an integer adjacent to $\lfloor\alpha \sqrt{L}\rfloor$ might be optimal. Therefore, not only the four values of step width as in Eq.~(\ref{prediction}) are tested, but also the two integers adjacent to the best of them. For a fair comparison the same number of trials is allowed in the random experiment. For 14 out of 17 networks the \MVM modularity value with $l$ as in Eq.~(\ref{prediction}) is higher or equal than for the corresponding random experiment (Table \ref{real-world-estimation}). Therefore, one can conclude that the empirical formula (\ref{prediction}) yields a step width that results in (close to) optimal modularity, and therefore only six runs of the \MVM algorithm are required.

\subsection{Quality of \MVM network partition}
Previously, the performance of the \MVM algorithm in optimizing modularity has been shown on 19 real-world networks \cite{Schuetz2007}. Here, an in-depth analysis of two examples provides evidence that the \MVM algorithm gathers vertices in groups that represent substructures (identified by other means) more accurately than the greedy algorithm.

\subsubsection{Metabolic network of \textit{E.~coli}}
The network of metabolic reactions in the bacterium \textit{E.~coli} is extracted from the KEGG database (data set ``Escherichia coli K-12 MG1655'') with additional refinement by Ma and Zeng \cite{Ma2003} and projected on the largest connected component. Furthermore, chains of vertices with degree 1 or 2 are reduced to one single vertex (to reduce spurious effects of unnaturally splitted chains). Each vertex is assigned to between zero and eight out of 11 metabolic pathways with an average of $1.51 \pm 0.99$. 

\begin{figure*}
 \centering
 \includegraphics[width=17cm]{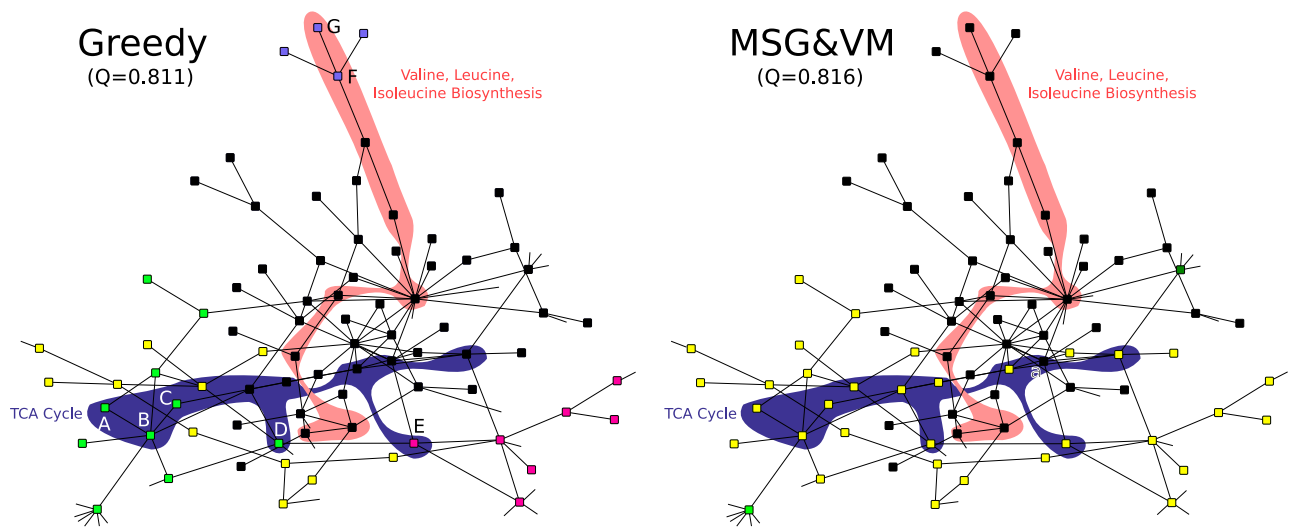}
 
\caption{\label{metabolic-ec}(Color online) Clusterization of the metabolic network of \textit{E.~coli} and accuracy of pathway identification. Two exemplary pathways as taken from the KEGG database \cite{Kanehisa2000, Kanehisa2002a} (pathways MAP00020 for ``TCA cycle'' and MAP00290 for ``Valine, Leucine, Isoleucine Biosynthesis'') are highlighted by the colored areas. An excerpt of the network is shown here while the full network is in the supplementary material \cite{suppl-mat}. The misassigned vertices are indicated by letters;  they are a=(S)-Malate for \MVMA, and for the original greedy: A=3-Carboxy-1-hydroxypropyl-ThPP, B=2-Oxoglutarate, C=Oxalosuccinate, D=Succinate, E=Fumarate, F=2-Oxoisovalerate, and G=Valine. }
\end{figure*}

Eleven communities are identical in the \MVM partition (which consists of 19 communities and has $Q=0.816$) and the partition obtained with the greedy algorithm (20 communities, $Q=0.811$). To assess the quality of pathway detection we employ the measure $P = \sum_{i} \frac{P_i}{N_i}$ (adopted from \cite{Muff2005}), with $P_i$ the number of vertex pairs in community $i$ that share at least one pathway and $N_i$ the number of pairs of vertices with equal community affiliation. The \MVM partition has $P_{\mbox{\tiny\MVM}}=0.60$,  which is better than the partition obtained with the original greedy algorithm ($P_\mathrm{greedy}=0.58$). The improved pathway identification is illustrated by  an excerpt of the network in Fig.~\ref{metabolic-ec} (vertices in the 11 modules which are identical in both partitions are removed for visibility reasons). Two central pathways (classification according to KEGG database) are highlighted by colored areas. In the \MVM solution the vertices of each pathway belong to separate modules except for ``(S)-Malate''. This metabolite has more connections to vertices assigned to the ``Amino Acid Metabolism'' than to those of the ``Carbohydrate Metabolism'' (the ``TCA cycle'' is associated to the latter) and thus, a separation is meaningful.  On the other hand, the metabolites misclassified by the original greedy algorithm are ``2-Oxo-glutarate'' (B), ``3-Carboxy-hydroxypropyl-ThPP''(A), and ``Oxalosuccinate''(C). The last two belong only to the ``TCA cycle'' pathway, whereas ``2-Oxo-glutarate'' is part of several pathways and therefore can also be attributed to other communities. Furthermore, the separation of the blue vertices in the ``Valine, Leucine, Isoleucine Biosynthesis'' pathway is peculiar as the overlapping pathway ``pantothenate and CoA biosynthesis'' is contracted to one vertex (the vertex right to ``F'' and ``G''). The metabolites ``F'' and ``G'' are the educts in the ``pantothenate and CoA biosynthesis'' pathway. If a unique assignment has to be made, an attribution to the ``Valine, Leucine, Isoleucine Biosynthesis'' pathway is more reasonable. The last differences of the greedy partition to the \MVM solution are ``Succinate'' (D) and ``Fumarate'' (E)  which are as ``(S)-Malate'' (a) part of multiple different metabolic processes and therefore may be attributed to multiple pathways. To summarize, of eight assignments differing between \MVM and original greedy algorithm (in the excerpt of the network shown in Fig.~\ref{metabolic-ec}), none was misplaced by the \MVM algorithm, whereas the greedy algorithm misplaced two  metabolites (two further examples of incomplete detection of  pathways by the original greedy algorithm are shown in the supplementary material \cite{suppl-mat}). 

\subsubsection{Network of words in titles of M.~Karplus' publications}
Martin Karplus is one of the most productive and most cited chemists (78091 citations as of July 3, 2008). As second example we analyze the community structure of the graph of words coappearing in the titles of the 719 publications (co)authored by M.~Karplus between 1947 and 2004 \cite{Schuetz2007,Karplus}. The words with highest degree in the five largest (according to number of words) communities are shown in Table \ref{Karplus}.
\begin{table*}
\begin{tabular}{llrlcl}\hline 
 & & &  & Number of titles \\[-3mm]
& & \multicolumn{2}{l}{\underline{Most frequent words}}  & with any of the\\[-3mm]  
Rank & Vertices & \phantom{ii}Degree & \phantom{M}Word & words in community & \multicolumn{1}{c}{Description} \\[-1mm] \hline
1& 220 & 407 & \phantom{M}Protein    &    442    &             molecular dynamics (of proteins)\\[-3mm]
& & 318 &\phantom{M}Simulation\\[-3mm]
& & 269 & \phantom{M}Molecular-dynamics\\[-1mm]
2& 184& 290 & \phantom{M}Structure  &   330   &             three-dimensional structures\\[-3mm]
& & 123 & \phantom{M}Peptide\\[-3mm]
 & & 97 & \phantom{M}Inhibitor\\[-1mm]
3& 162 & 269 & \phantom{M}Model & 335 & molecular modelling, \\[-3mm]
& & 178 & \phantom{M}Energy & &  molecular mechanics \\[-3mm]
& & 169 & \phantom{M}Function \\[-1mm]
4& 162& 159 & \phantom{M}Molecule   &   306      &          quantum mechanics, \\[-3mm]
& & 154 & \phantom{M}Free-energy  & &  free-energy calculation\\[-3mm]
& & 144 & \phantom{M}Potential\\[-1mm]
5& 116& 212 & \phantom{M}Reaction   &    205  &               chemical reaction, kinetics, \\[-3mm]
& & 154 & \phantom{M}Solution  & & and solvation\\[-3mm]
& & 101 & \phantom{M}Solvation\\[-1mm]
\hline 
\end{tabular}
\caption{\label{Karplus}The five largest communities as identified by the \MVM algorithm in the network of words in the titles of M.~Karplus' papers. These five communities account for 81\% of the vertices in the network. Unspecific words (e.g., ``study'' and ``theory'' with degree 291 and 234, respectively) were taken into account for the clusterization, but are not listed in this table.}
\end{table*}

\hyphenation{hydro-lysis}
The following two examples provide evidence for
the superiority of the \MVM partition
(11 communities, $Q=0.316$) with respect to 
the partition obtained by the original greedy algorithm
(18 communities, $Q=0.264$).
The words ``reaction'' (degree 212), ``hydrolysis'' (73),
``rate'' (69), ``enzyme'' (57), ``catalysis'' (54), and ``kinetics'' (54)
are appropriately grouped in a single community by the former, while they are
spread in the four largest (according to the number of words)
communities by the latter.
Another example of superiority of the \MVM partition
is the community with the words ``molecule'' (159), ``atom'' (91),
and ``bond'' (87), which are spread over 
 the three largest communities by the greedy algorithm.
These two examples show that the main advantage
of the \MVM algorithm is that the simultaneous
emergence of several communities hinders the
spurious coalescence into few large communities
observed for the original greedy algorithm.

\section{Conclusions}
The performance of the MSG procedure, a multistep extension of the greedy algorithm, was analyzed on 1100 computer-generated networks of heterogeneous size and degree distributions and 17 real-world networks. Several powers of topological properties (e.g., average degree, clustering coefficient etc.) were tested as prediction formulas for the optimal step width $l$. The empirical formula $l=\lfloor \alpha \sqrt{L}\rfloor$ ($L$ total edge weight; $\alpha=0.25, 0.5, 0.75, 1$) outperforms all others and yields a higher modularity value than a random picking of the step width for 85.5\% of the computer-generated networks  and 14 of 17 real-world examples. For these 14 real-world networks, the modularity optimized by \MVM algorithm using only six values of $l$ ($l_1=\lfloor 0.25 \sqrt{L}\rfloor, l_2= \lfloor 0.5 \sqrt{L}\rfloor, l_3=\lfloor 0.75 \sqrt{L} \rfloor, l_4=\lfloor 1.0 \sqrt{L}\rfloor$, and $l_{5,6}=l_{\rm max}\pm 1$ with $l_{\rm max}$ the step width among $l_{1,\ldots,4}$ that yields the highest modularity) is larger than 99\% of the highest value achievable by exhaustive testing of all step widths (i.e., $1\leq l \leq L$). This deviation is on the order of the fluctuations observed when the parsing order of the vertices is changed. In addition, for 92.6\% of the computer-generated and 13 of 17 real-world networks the optimal value of the step width is smaller than $1.5 \sqrt{L}$. 

To assess the quality of the community identification two real-world examples (the network of metabolic reactions in \textit{E.~coli} and the graph of coappearing words in titles of publications coauthored by M.~Karplus) were examined in-depth and the modular structure obtained from the application of the \MVM and greedy algorithms was compared. For the metabolic network the original greedy algorithm splits two exemplary pathways (``TCA cycle'' and ``Valine, Leucine, Isoleucine Biosynthesis'') in multiple parts with seven misplaced vertices. Two of these vertices are not part of another pathway and therefore are wrongly assigned by the original greedy algorithm. For the \MVM solution only one metabolite is misplaced which can be attributed to the three pathways in which this metabolite is involved. Furthermore, an objective criterion (the conditional probability that two vertices in the same module share at least one pathway) supports these exemplary observations. For the ``M.~Karplus'' network the partition obtained by the original greedy algorithm has three very large modules in which words of distinct research fields are inappropriately mixed. Moreover, subsets of words belonging to the same topic are erroneously split (e.g., ``atom'', ``molecule'', and ``bond'' are split in the three largest modules). On the other hand, the \MVM procedure more accurately groups subsets of words belonging to individual research topics.

In conclusion, the \MVM algorithm is one of the fastest and most accurate procedures for modularity optimization currently available because \rc{it scales as $O(N \log^2 N )$ for a sparse network ($N$ the number of vertices) \cite{Schuetz2007}. Therefore, a single run is faster than previously published approaches \cite{Danon2005}}, and only six independent runs are required using Eq.~(\ref{prediction}) to determine the step width \cite{Karplus}.

\section{Acknowledgments}
We thank Stefanie Muff for comments on the manuscript. Christian Bolliger, Thorsten Steenbock and Dr.~Alexander Godknecht are acknowledged for maintaining the Matterhorn cluster where most of the presented parameter studies were performed. For providing the data sets we are thankful to Dr.~A.~Arenas, Dr.~A.~L.~Barab\'asi, Dr.~P.~Gleiser, Dr.~H.~Ma and Dr. ~M.~E.~J.~Newman  \cite{Karplus}. This work was supported by a Swiss National Science Foundation grant to A.C.

\end{document}